\documentclass[a4paper,10pt]{article}

\title{Domain Wall Displacement Detection Technology Research Report\footnote{Chinese version of this paper is finished Feb.10 2003. And it was translate to English in Aug.17 2005}}

\author{Ran Ren\thanks{Department of Telecommunications Engineering, Nanjing Univ. of P. \& T., Nanjing, Jiangsu 210046, P.R.China. Email: ranren@ieee.org}}

\date{Aug.17 2005}

\begin{document}

\maketitle

\begin{abstract}
This article introduce a new data storage method called DWDD(Domain Wall
Displacement Detection) and tell you why it succeed.
\end{abstract}


\section{History}
Nowadays, media communication is greatly developed. By the development of media communication, a large-capacity, high-density disks which also can be erased repeatedly are raring acquired.\\

By the time, SONY corporation in Japan was released their newly developed, next generation Minidisc player, Hi-MD on January 7,2004 at Las Vegas, USA. The invention of Hi-MD series Minidisc player symbolize that Minidisc gains advantages in the competition with MP3 player, especially the Apple's iPod. Hi-MD have larger disc storage spaces. And DWDD(Domain Wall Displacement Detection) Technology is the main technology of next generation of MD disc
which's storage space is largely increased. So it is critical for us to know and research this technology.\\

In fact, scientists and engineers have done lots of jobs in order to largely increase optical magnetic disc's storage space and density. For instance, the famous magnetic field modulation technology use laser beam of mask. But with the tend of increasing storage space, how to use laser beams to read a narrower masks is the bottleneck of the development of the technology.(Because the width of laser beam are certain) This is also the point that world-wide scientists trying to solve. The advantage of this technology is certain,if the scientists really develop that kind of technology, the width of the masks can be infinitely small theoretically. In the other words,if a technology of reading tiny masks be developed, the density of the optical magnetic disc can be infinitely high theoretically. Many corporations have solutions about it. For example, magnetic super-resolution technology and magnetic domain enlargement detection
technology. But the insolvability of these technologies are mush dependent on the area of temperature distributing on the disc,so the formation of these area is due to the frequency of the laser beam.\\

The difficulties of these kinds of technologies are evident, in order to meet the compatibility with previous series of products, the frequency of the existing laser beams is unable to change, in the other word, the wavelength of the laser beam is certain. So the difficulties are seen, how to use a longer wavelength laser beam to read a narrower width of information masks? In physics, that seems to be mission impossible.\\

But, the mankind intelligence is great, Sony and an other Japanese corporation Canon announced a kind of solution in 2000. This technology has a patent in United States Patent and Trade Mark Organizations(USPTO)\cite{USPatent}. This is what we call Domain Wall Displacement Detection(DWDD) technology.\\

\section{Approaches of data storage on optical magnetic discs}
The Data in optical magnetic discs is not purely stored as ``data flow", instead, the data is divided into many tiny masks, then these tiny masks are stored in disc at certain mask intervals. It is easy to discover that the density of the disc is constrained by the wavelength of the laser beam, so the width of masks only can be as narrow as the wavelength of the laser beam. If the width of masks is narrower, the laser beam will not read them. So, it looks that if you want to increase the density of the disc, the only solution is adopt the narrower wavelength laser beams, it is widely use blue LD to
achieve this. But to optical magnetic discs like MD, change the laser beams means not compatible with the previous products. So the approach of change the laser beam of increase the density of the disc is not feasible.\\

\section{What is DWDD}
The DWDD technology use a physics phenomena called Domain Wall Displacement, which implements shorten the masks but don not need to change the laser beam. With such technology, the density of the disc can be increased dramatically.

\section{The infrastructure of DWDD}
The innovative part of DWDD technology is the three-tier infrastructure of reading mechanism. The layer in the bottom is called memory layer, the layer
nearest to the laser beam is named displacement layer, the middle layer is switching layer. Not only structural there are three different, there are three different materials in the three layers. We know, Minidisc belongs to optical magnetic disc, when data is write into optical magnetic disc, the optical head and the magnetic head are all used. How it use two heads? Physically, every materials have its own temperature of losing magnetism. In an other word, when the material is heated to the certain temperature, the magnetism of the material will be lost and can be magnetized over again. This temperature of material is called Curie temperature.\\

In the three layers, memory layer is make up of a perpendicular magnetizing film which is magnetized by two upright ordinate magnetic fields, the reluctance of its magnetic wall is so strong, it means the existing state of the magnetic wall is hard to change, but the displace force is so strong.
The Curie temperature of the material of switching layer is lowest, so when laser beam irradiate on the disc, the switching layer lose magnetism first. This induce the exchange coupling force between each layer lost. Exchange coupling force is one of the forces that maintains the size of the tiny
masks which is displaced from the fact storage area(memory layer) to displacement layer. When the exchange coupling force is disappeared due to the disappearance of the magnetism of the switching layer, the temperature of domain wall around that tiny masks rises, the domain wall energy reduces. So domain wall on longer has the force that ties the tiny masks in its original small space, then, the tiny record masks enlarged.\\

Since the enlargement of the tiny record masks is due to disappearance of exchange coupling force, and only when the temperature meets the Curie temperature of the switching layer material, in an other word, this can only occurred when the laser beam is irradiated on the surface of the disc. The
optical magnetic discs' circumgyratetion velocity is fairly large, so the displacement can be seen as a instantaneous action, and it is only occurred when data is been read from the disc. The situation of the displacement is seems that data is passing a magnifier. By this, we can use existing laser beam to read data even if the density of the disc is extremely high.\\

\section{Data reading in DWDD}
There are three steps of data reading in the optical magnetic disc which adopted DWDD technology.
\begin{itemize}
\item Let the laser beam irradiated to the displacement layer and induce the asymmetry of temperature distribution in the switching layer, so the temperature of some areas id higher then Curie temperature. So in the direction of the movement of laser beam, the magnetic wall displacement
occurs. Use this approach to acquire the changing reproduced signal on the polarization plane of the reflection beam of the laser beam.
\item Use the signal produced in step 1 to produce reincarnate signal, then use the differential signal to detect whether or not the domain wall is displaced, then produce the detection signal.
\item Use the detection signal produced in step 2 to detect data, finally read the data out.
\end{itemize}

\section{Future of DWDD}
The mass storage medium that implements DWDD technology can be competent for storage of long-time video data. The technology can not only be brought into play in AV(Audio and Video) equipments, but also be applied to PC(Personal Computer) and home network recording equipments. In the future, the compaction of DWDD is feasible, so DWDD can also by used in wireless devices. What is more, using this approach, existing optical and magnetic heads can be
used without any modifications. So it also reduce the cost, more user can take it. We believe, with so many advantages, DWDD will be a new standard in the new multimedia era.\\

\end{document}